\documentclass[12pt]{article}

\usepackage{graphicx}
\usepackage{a4}

\usepackage{fancybox}
\usepackage{epsfig}
\usepackage{color}

\usepackage{amsfonts}
\usepackage{mathrsfs}
\usepackage{amssymb}
\usepackage{amsmath}

\usepackage{dcolumn}
\usepackage{bm}

\def\pa{\partial}

\def\al{\alpha}
\def\be{\beta}

\def\th{\theta}

\def\om{\omega}

\def\La{\Lambda}

\def\Om{\Omega}

\newcommand{\ben}{\begin{equation}}
\newcommand{\een}{\end{equation}}
\newcommand{\bea}{\begin{eqnarray}}
\newcommand{\eea}{\end{eqnarray}}
\newcommand{\ba}{\begin{array}}
\newcommand{\ea}{\end{array}}
\newcommand{\bit}{\begin{itemize}}
\newcommand{\eit}{\end{itemize}}

\newcommand{\vs}[1]{\vspace{#1 mm}}

\newcommand{\dsl}{\pa \kern-0.5em /}

\begin{document}

\topmargin 0pt \oddsidemargin 0mm

\vspace{2mm}

\begin{center}

{\Large \bf Expanded solutions of force-free electrodynamics on
general Kerr black holes}

\vs{10}

 {\large Huiquan Li \footnote{E-mail: lhq@ynao.ac.cn} and Jiancheng Wang}

\vspace{6mm}

{\em

Yunnan Observatories, Chinese Academy of Sciences, \\
650216 Kunming, China

Key Laboratory for the Structure and Evolution of Celestial Objects,
\\ Chinese Academy of Sciences, 650216 Kunming, China

Center for Astronomical Mega-Science, Chinese Academy of Sciences,
\\ 100012 Beijing, China}

\end{center}

\vs{9}

\begin{abstract}
In this work, expanded solutions of force-free magnetospheres on
general Kerr black holes are derived through a radial distance
expansion method. From the regular conditions both at the horizon
and at spatial infinity, two previously known asymptotical solutions
(one of them is actually an exact solution) are identified as the
only solutions that satisfy the same conditions at the two
boundaries. Taking them as initial conditions at the boundaries,
expanded solutions up to the first few orders are derived by solving
the stream equation order by order. It is shown that our extension
of the exact solution can (partially) cure the problems of the
solution: it leads to magnetic domination and a mostly timelike
current for restricted parameters.
\end{abstract}



\section{Introduction}
\label{sec:introduction}

Black hole magnetospheres are believed to play essential roles in
many high-energy astronomical objects. In the popular
Blandford-Znajek model \cite{Blandford:1977ds,Znajek:1977unknown},
energy can be extracted from a rotating black hole via a stationary
and force-free magnetosphere on it to eject dipole relativistic
jets, which may account for most of the high-energy phenomena in
active galactic nuclei, gamma-ray bursts and microquasars.

In the simplest configuration, the force-free magnetosphere is well
described by the clean and precise electrodynamics on a Kerr black
hole. However, the present understanding of such a system strongly
relies on numerical simulations. We have very few options in
analytical approaches to date.

One of the approaches is the perturbation method first given in the
original work of Blandford and Znajek \cite{Blandford:1977ds}. Based
on the split monopole and paraboloidal solutions on a non-rotating
black hole, analytical solutions on a slow-rotating black hole are
derived by expanding the functions and stream equation to leading
orders of the spin parameter. So the solutions apply to slowly
rotating black holes. To get analytical properties of magnetospheres
on rapidly rotating black holes, which may be more interesting to
us, we need to calculate higher order corrections. But it seems
uneasy to do so \cite{Tanabe:2008wm}. Recently, the solution up to
the fourth order has been derived \cite{Pan:2015haa}.

Solutions that go beyond the slow-rotation limit in the perturbation
approach can be obtained in some limited regions. In the work
\cite{Menon:2005va,Menon:2005mg} of Menon and Dermer (MD),
asymptotic solutions (and their generalisation
\cite{Brennan:2013jla}) were derived in regions far away from the
horizon. The solutions can apply to black holes with general angular
momentum. But these solutions are radial distance independent. In
particular, one of the solutions is even the only known exact
solution so far that can solve the full stream equation, which
leaves it very interesting. However, the current for this solution
is along the infalling principle null geodesic. This means that
charged particles must move at the speed of light, which is not
allowed. Besides, the electromagnetic fields from the solution are
also null. A relieving method is given by the authors in
\cite{Menon:2011zu,Menon:2014vqa}. The lightlike current is
artificially decomposed into a linear combination of two timelike
currents with opposite charges.

On the other hand, in past years, exact solutions on extremely fast
rotating black holes were obtained by focusing on the near-horizon
region
\cite{Lupsasca:2014pfa,Lupsasca:2014hua,Li:2014bta,Zhang:2014pla,
Compere:2015pja}. But, a smooth connection between these near- and
far-region solutions is lacking.

In this work, we consider a different expansion method other than
the one in the traditional perturbation approach. In terms of the
boundary conditions of a magnetosphere, we expand the functions and
stream equation in series of the radial distance, instead of the
spin parameter. Analytical solutions that depend on both poloidal
coordinates can be derived order by order following a precise
procedure. So this approach hopefully can help us extend solutions
in far region to the ones in the near region. Moreover, this
provides a method to generalise the MD exact solution and relax its
problems of null current and electromagnetic fields.

The paper is organised as follows. In Section \ref{sec:streameq},
the stream equation of a force-free magnetosphere is instructed and
presented. In Section \ref{sec:bc}, we show the boundary conditions
at the horizons and at infinity, which can be determined from the
stream equation. Two special cases of the boundary conditions lead
to the previously known asymptotic solutions. In Section
\ref{sec:expmethod}, the expansion forms and solving procedure of
the stream equation are introduced in terms of the boundary
conditions. Examples of solutions are derived and analysed in
Section \ref{sec:solutions}. Then we summarise in the last section.

\section{The stream equation}
\label{sec:streameq}

On the Boyer-Lindquist coordinates, a Kerr black hole is depicted by
the metric:
\begin{equation}\label{e:Kerr}
 ds^2=-\La^2 dt^2+\frac{\rho^2}{\triangle}dr^2+
\rho^2 d\th^2+\varpi^2(d\phi-\om dt)^2,
\end{equation}
where
\begin{equation}
 \rho^2=r^2+a^2\cos^2\th, \textrm{ }\textrm{ }\textrm{ }
a=\frac{J}{M}, \nonumber
\end{equation}
\begin{equation}
 \La^2=\frac{\rho^2\triangle}{A}, \textrm{ }\textrm{ }\textrm{ }
\varpi^2=\frac{A\sin^2\th}{\rho^2}, \textrm{ }\textrm{ }\textrm{ }
\om=\frac{2Mar}{A}, \nonumber
\end{equation}
\begin{equation}
 \triangle=(r-r_+)(r-r_-), \textrm{ }\textrm{ }\textrm{ }
r_\pm=M\pm\sqrt{M^2-a^2}, \nonumber
\end{equation}
\begin{equation}
 A=(r^2+a^2)^2-a^2\triangle\sin^2\th=2Mr(r^2+a^2)+\triangle\rho^2.
\nonumber
\end{equation}
The spin parameter $a$ measures the angular momentum $J$ per unit
mass $M$ of the black hole. The inner and outer horizons are located
at $r=r_-$ and $r=r_+$ respectively. From the above relations, the
velocities of the black hole at the horizons are:
\begin{equation}
 \om_\pm\equiv \frac{a}{r_\pm^2+a^2}.
\end{equation}

In the $3+1$ split formulation \cite{MacDonald:1982zz}, the
four-dimensional spacetime (\ref{e:Kerr}) is replaced by a
three-dimensional absolute space and a universal time coordinate.
The electrodynamics on a Kerr black hole can be equivalently dealt
with on the following absolute space:
\begin{equation}\label{e:absolutesp}
 ds_A^2=\frac{\rho^2}{\triangle}dr^2+
\rho^2 d\th^2+\varpi^2d\phi^2.
\end{equation}
The four-dimensional quantities and equations are split accordingly.
The quantities we deal with on the absolute space are measured by
the so-called zero-angular momentum observers (ZAMOs). From the
inverse metric, the unit basis vectors are given by
\begin{equation}\label{e:ffcon}
 \textbf{e}_{\hat{r}}=\sqrt{\frac{\triangle}{\rho^2}}\pa_r,
\textrm{ }\textrm{ }\textrm{ }
\textbf{e}_{\hat{\th}}=\frac{1}{\sqrt{\rho^2}}\pa_\th, \textrm{
}\textrm{ }\textrm{ }
\textbf{e}_{\hat{\phi}}=\frac{\sqrt{\rho^2}}{\sqrt{A}\sin\th}\pa_\phi.
\end{equation}

The Kerr spacetime has Killing vectors along the time and along the
toroidal directions. For simplicity, we consider the stationary and
axisymmetric case of electrodynamics on the spacetime. The relevant
inhomogeneous Maxwell's equations relate the electromagnetic fields
to the electric charge and current densities ($\rho_e,\textbf{j}$):
\begin{equation}\label{e:inMaxeqs1}
 \nabla\cdot\textbf{E}=4\pi\rho_e,
\end{equation}
\begin{equation}\label{e:inMaxeqs1}
 \nabla\times(\La\textbf{B})=4\pi\La\textbf{j}-\varpi(\textbf{E}
\cdot\nabla\om)\textbf{e}_{\hat{\phi}}.
\end{equation}
Throughout the paper, the operator $\nabla$ denotes the covariant
derivative associated with the 3-dimensional spatial dimensions
(\ref{e:absolutesp}). The homogeneous Maxwell's equations tell us
that the electromagnetic fields can be expressed as the gauge
potentials ($A_0,\textbf{A}$):
\begin{equation}\label{e:Egen}
 \textbf{E}=\frac{1}{\La}(\nabla A_0+\om\nabla A_\phi),
\end{equation}
\begin{equation}\label{e:Bgen}
 \textbf{B}=\nabla\times\textbf{A}.
\end{equation}

We ignore the dynamics of plasma and impose the force-free condition
\begin{equation}\label{e:ffcon}
 \rho_e\textbf{E}+\textbf{j}\times\textbf{B}=0,
\end{equation}
which automatically satisfies
\begin{equation}\label{e:degcon}
 \textbf{j}\cdot\textbf{E}=0, \textrm{ }\textrm{ }\textrm{ }
\textbf{E}\cdot\textbf{B}=0.
\end{equation}
Under the conditions, the electrodynamics is described by three
correlated functions: the flux $\psi=2\pi A_\phi$ and the total
electric current $I(\psi)$ flowing through the area enclosed by an
axisymmetric loop, and the angular velocity of the electromagnetic
field lines $\Om(\psi)=-dA_0/dA_\phi$ on the loop.

The electromagnetic fields read
\begin{equation}\label{e:E}
 \textbf{E}=-\frac{\Om-\om}{2\pi\La\sqrt{\rho^2}}\left(\sqrt{\triangle}
\pa_r\psi\textbf{e}_{\hat{r}}+\pa_\th\psi\textbf{e}_{\hat{\th}}\right),
\end{equation}
\begin{equation}\label{e:B}
 \textbf{B}=\frac{1}{2\pi\sqrt{A}\sin\th}\left(\pa_\th\psi
\textbf{e}_{\hat{r}}-\sqrt{\triangle}\pa_r\psi\textbf{e}_{\hat{\th}}
+\frac{4\pi I\sqrt{\rho^2}}{\La}\textbf{e}_{\hat{\phi}}\right).
\end{equation}
The charge and current densities are respectively
\begin{equation}\label{e:charge}
 \rho_e=-\frac{1}{8\pi^2}\nabla\cdot\left(\frac{\Om-\om}{\La}
\nabla\psi\right),
\end{equation}
\begin{equation}\label{e:current}
 \textbf{j}=\frac{1}{\La}[\rho_e\varpi(\Om-\om)
\textbf{e}_{\hat{\phi}}+I'\textbf{B}],
\end{equation}
where the prime stands for the derivative with respect to $\psi$.
Note that the total current $I$ is defined to flow upwards, with
opposite sign to that defined in the original paper
\cite{MacDonald:1982zz}.

From the above equations and expressions, we will find that the
force-free electrodynamics on a Kerr spacetime can be described by
the following unique stream equation \cite{MacDonald:1982zz}
\begin{equation}\label{e:GSeq}
 \nabla\cdot\left\{\frac{\La}{\varpi^2}\left[1-
\frac{(\Om-\om)^2\varpi^2}{\La^2}\right]\nabla\psi
\right\}+\frac{\Om-\om}{\La}\Om'(\nabla\psi)^2+
\frac{16\pi^2}{\La\varpi^2}II'=0.
\end{equation}

For the electromagnetic system, the poloidal components of the
energy and angular momentum flux densities from the hole are given
by:
\begin{equation}\label{e:r-enmom}
 \mathcal{E}^r=\Om\mathcal{L}^r=-\Om\frac{I\pa_\th\psi}
{2\pi\sin\th\rho^2},
\end{equation}
\begin{equation}\label{e:th-enmom}
 \mathcal{E}^\th=\Om\mathcal{L}^\th=\Om\frac{I\pa_r\psi}
{2\pi\sin\th\rho^2}.
\end{equation}

\section{Boundary behaviours}
\label{sec:bc}

In regions that are accessible to us, the differential equation
(\ref{e:GSeq}) has two boundaries: one at the horizon and the other
at spatial infinity (if the force-free region extends far away from
the outer horizon). In some sense, the two boundaries have similar
behaviours and features, e.g., they both attain the following
radiation condition \cite{Nathanail:2014aua}:
\begin{equation}\label{e:radcon}
 E^\th=\pm B^\phi,
\end{equation}
for the electromagnetic fields (\ref{e:E}) and (\ref{e:B}) as they
are approached. The condition can be obtained directly from the
stream equation (\ref{e:GSeq}). To make this clear, we re-express
the stream equation as the following form
\begin{eqnarray}\label{e:reGSeq}
 \frac{\triangle}{A}\left\{
\left[\frac{A^2\sin^2\th(\Om-\om)^2}{\rho^4}-\triangle
\right]\pa_r^2\psi+\frac{A^2\sin^2\th(\Om-\om)}{\rho^4}
\pa_r\Om\pa_r\psi\right.
\nonumber \\
+\frac{2A\sin^2\th}{\rho^2}\left[r\Om^2-
\frac{Ma^2\sin^2\th(r^2-a^2\cos^2\th)\left(
\Om-\Om_N\right)^2}{\rho^4}\right]\pa_r\psi
\\
\left.-\pa_\th^2\psi+\frac{1}{2}\left[\frac{\triangle}{A}+
\frac{A\sin^2\th\left(\Om^2-\Om_N^2\right)}{\rho^4}\right]
\pa_\th\rho^2\pa_\th\psi\right\}
\nonumber \\
+\frac{\sqrt{A}\sin\th(\Om-\om)}{\rho^2}\pa_\th
\frac{\sqrt{A}\sin\th(\Om-\om)\pa_\th\psi}{\rho^2} -16\pi^2II'=0,
\nonumber
\end{eqnarray}
where $\Om_N=1/(a\sin^2\th)$.

\subsection{Conditions at horizons}

From the equation (\ref{e:reGSeq}), we can see that only the last
line of the equation remains at the horizons $r=r_\pm$ or
$\triangle=0$:
\begin{eqnarray}\label{e:horeq}
 16\pi^2I\frac{\pa }{\pa\psi}I=\frac{\sqrt{A}\sin\th(\Om-\om)}
{\rho^2}\pa_\th\frac{\sqrt{A}\sin\th(\Om-\om)\pa_\th\psi}{\rho^2}
\nonumber \\
=\frac{\sqrt{A}\sin\th(\Om-\om)\pa_\th\psi}{\rho^2}\frac{\pa}
{\pa\psi}\frac{\sqrt{A}\sin\th(\Om-\om)\pa_\th\psi}{\rho^2}.
\end{eqnarray}
Approaching the event horizon, $\psi$ is a function only dependent
on $\th$ \cite{MacDonald:1982zz}. This structure in the stream
equation is also found in the near-horizon treatments of
magnetospheres on near-extreme Kerr black holes \cite{Li:2014bta}.

From the above relation at the horizons, we have
\begin{equation}\label{e:horcon}
 r=r_\pm: \textrm{ }\textrm{ }\textrm{ }
I^2=C_\pm+\left[\frac{\sqrt{A}\sin\th(\Om-\om)}{4\pi\rho^2}
\pa_\th\psi\right]^2,
\end{equation}
where $C_\pm$ are constants. So we can conclude that any solution
$\psi$ (with any given correlated functions $\Om_F(\psi)$ and
$I(\psi)$) satisfying the stream equation (\ref{e:reGSeq}) will
always satisfy the condition (\ref{e:horcon}), only if the sum of
the terms within the brace in Eq.\ (\ref{e:reGSeq}) are non-singular
compared with the rest terms at the horizons .

It is easy to find that the Znajek boundary condition can be
obtained by setting the special value:
\begin{equation}\label{e:Zbd}
 C_\pm=0.
\end{equation}
When the positive sign is chosen, the conditions (\ref{e:horcon}) at
the horizons read
\begin{equation}\label{e:Z+bd}
 I_+
=\frac{Mr_+ \sin\th(\Om_+-\om_+)}{2\pi\rho_+^2}\pa_\th\psi_+,
\end{equation}
\begin{equation}\label{e:Z-bd}
 I_-
=\frac{Mr_- \sin\th(\Om_--\om_-)}{2\pi\rho_-^2}\pa_\th\psi_-,
\end{equation}
where $\psi_\pm=\psi(r_\pm)$, $\Om_\pm=\Om(\psi(r_\pm))$,
$I_\pm=I(\psi(r_\pm))$ and $\rho_\pm^2=r_\pm^2+a^2\cos^2\th$. The
former is exactly the Znajek regularity condition at the outer
horizon \cite{Znajek:1977unknown}, which corresponds to the positive
sign case of the radiation condition (\ref{e:radcon}):
$E^\th=B^\phi$. The positive sign is chosen because this means
current flow is directed outwards for $0<\Om_+<\om_+$
\cite{MacDonald:1982zz,Ruiz:2012te}, which leads to energy and
angular momentum extraction from the hole across the event horizon,
as implied by Eq.\ (\ref{e:r-enmom}).

\subsection{Condition at spatial infinity}

Let us now turn to the behaviours at spatial infinity. As shown in
Eq.\ (\ref{e:r-enmom}) and (\ref{e:th-enmom}), the energy and
momentum extraction rates are different by the angular velocity
$\Om$. Since the energy and momentum extracted from the hole must be
finite at spatial infinity, $\Om$ should be independent of $r$ at
infinity:
\begin{equation}\label{e:Om-infcon}
 \Om(r,\th)\rightarrow\Om_0(\th) \textrm{ }\textrm{ } \textrm{ as }
\textrm{ }\textrm{ } r\rightarrow\infty,
\end{equation}
as noticed in \cite{Menon:2005va,Menon:2005mg}. Further, since
$I(\Om)$ and $\psi(\Om)$ are functions of $\Om$, then the associated
functions $I_0$ and $\psi_0$  at infinity should be functions of
$\Om_0$ as well:
\begin{equation}\label{e:psi-I-infcon}
 r\rightarrow\infty: \textrm{ }\textrm{ }\textrm{ }
\psi(\Om)\rightarrow\psi_0(\Om_0(\th)), \textrm{ }\textrm{ }\textrm{
} I(\Om)\rightarrow I_0(\Om_0(\th)).
\end{equation}
That is, all three correlated functions should be independent of $r$
at infinity if one is. This is quite similar to the situation at the
outer horizon, where the functions are also only dependent on $\th$.

In the present work, we consider the case that $\Om_0$ is not a
constant or other trivial functions of $\th$  (so do $\psi_0$ and
$I_0$). With these asymptotic conditions (\ref{e:Om-infcon}) and
(\ref{e:psi-I-infcon}), we can find that the stream equation
(\ref{e:reGSeq}) becomes the following simple form at infinity:
\begin{equation}\label{e:}
 16\pi^2I_0\frac{\pa I_0}{\pa\psi_0}
=\sin\th\Om_0\pa_\th(\sin\th\Om_0\pa_\th\psi_0).
\end{equation}
Similarly, we have
\begin{equation}\label{e:}
 I_0^2=C_0+\frac{1}{16\pi^2}(\sin\th\Om_0\pa_\th\psi_0)^2,
\end{equation}
where $C_0$ is a constant. So any solutions satisfying the boundary
conditions (\ref{e:Om-infcon}) and (\ref{e:psi-I-infcon}) must
satisfy this relation. Here, we also choose the special case
$C_0=0$, for which the above relation reads
\begin{equation}\label{e:I0exp}
 I_0=-\frac{1}{4\pi}\sin\th\Om_0\pa_\th\psi_0.
\end{equation}
Here, the negative sign is chosen when $\Om_+\leq\om_+$, which
guarantees outflow of energy by inserting the current
(\ref{e:I0exp}) into Eq.\ (\ref{e:reGSeq}). This also corresponds to
the positive sign case of the radiation condition (\ref{e:radcon}).
Note that, when $\Om_+>\om_+$, we need to choose positive sign in
the equation (\ref{e:I0exp}) (corresponding to the minus sign case
of the radiation condition: $E^\th=-B^\phi$), which leads to influx
of energy at spatial infinity. The reason is that the direction of
energy can not reverse on a field line \cite{Blandford:1977ds}. If
the energy inflows across the event horizon for $\Om_+>\om_+$, we
should also have influx at infinity.

\subsection{The cases with identical boundary conditions}
\label{sec:spebc}

As in usual second-order differential equations, a set of solutions
can be defined by constraining appropriate conditions on the two
boundaries. On the other hand,  as we stated above, the behaviours
are similar at the two boundaries: the functions are purely
$\th$-dependent and satisfy the radiation condition
(\ref{e:radcon}). So it is natural to consider the special cases
that the conditions on the two boundaries are identical.

Generalising the condition (\ref{e:I0exp}) to include the positive
sign case, we can express the boundary condition at infinity as
\begin{equation}\label{e:genI0con}
 \pm\frac{4\pi}{\sin\th}=\Om_0\frac{\pa_\th\psi_0}{I_0}.
\end{equation}
On the other hand, the Znajek boundary condition (\ref{e:Z+bd}) can
be re-expressed as
\begin{equation}\label{e:reZ+bd}
 4\pi\left(\frac{1}{\sin\th}-a\sin\th\om_+\right)=(\Om_+
-\om_+)\frac{\pa_\th\psi_+}{I_+}.
\end{equation}
Now we consider the special case that the functions satisfy the same
boundary conditions at the horizon and at
infinity\footnote{Actually, the latter two stringent conditions can
be simply replaced by the unique one
$\pa_\th\psi_0/I_0=\pa_\th\psi_+/I_+$ when the relations between
$I$, $\psi$ and $\Om$ are not necessarily the same at the two
boundaries.}:
\begin{equation}\label{e:}
 \Om_0=\Om_+, \textrm{ }\textrm{ }\textrm{ }
\psi_0=\psi_+, \textrm{ }\textrm{ }\textrm{ } I_0=I_+.
\end{equation}

(1) If we choose the positive sign in Eq.\ (\ref{e:genI0con}), we
can have $\Om_+\pa_\th\psi_+/I_+=4\pi/\sin\th$ and
$\pa_\th\psi_+/I_+=4\pi a\sin\th$ by comparing the equations
(\ref{e:genI0con}) and (\ref{e:reZ+bd}). From the factor difference
between them, we have:
\begin{equation}\label{e:OmN}
 \Om_+(\th)=\Om_0(\th)=\Om_N\equiv\frac{a}{2Mr_+-\rho^2_+}
=\frac{1}{a\sin^2\th}.
\end{equation}
As expected, the angular velocity is larger than the one of the
black hole. That is why we have chosen positive sign in the
condition (\ref{e:genI0con}), as stated in the previous subsection.

(2) If we choose the negative sign in Eq.\ (\ref{e:genI0con}), we
have: $\Om_+\pa_\th\psi_+/I_+=-4\pi/\sin\th$ and
$\pa_\th\psi_+/I_+=-4\pi(2Mr_++\rho_+^2)/(a\sin\th)$, which leads to
\begin{equation}\label{e:OmP}
 \Om_+(\th)=\Om_0(\th)=\Om_P\equiv\frac{a}{2Mr_++\rho^2_+}.
\end{equation}

The two solutions (\ref{e:OmN}) and (\ref{e:OmP}) at boundaries are
exactly the same as the asymptotical solutions found in
\cite{Menon:2005va,Menon:2005mg} (the MD solutions). The first
solution is even the only known exact solution to date that can
solve the full stream equation. In deriving the above solutions, the
functions $I$ and $\psi$ are identical but not specified at the
boundaries.

In what follows, we only take them as initial values at the two
identical boundaries, instead of asymptotical solutions, to explore
analytical solutions that are $(r,\th)$-dependent in between the
boundaries.

\section{The expansion method}
\label{sec:expmethod}

In terms of the boundary properties, we may derive solutions to the
stream equation by expanding the functions in series of the radial
distance $r$, as done in Appendix for Schwarzschild black hole case.
If $\Om_0$, $\psi_0$ and $I_0$ are all nontrivial functions of $\th$
(i.e., not zero or constant), we can take the three correlated
functions as the following general expanded forms:
\begin{eqnarray}\label{e:exp}
 \Om=\sum_{n=0}^{\infty}\Om_{-n}(\th)r^{-n}, \nonumber \\
\psi=\sum_{n=0}^{\infty}\psi_{-n}(\th)r^{-n},  \\
I=\sum_{n=0}^{\infty}I_{-n}(\th)r^{-n}. \nonumber
\end{eqnarray}
We assume these expanded forms to be valid in all force-free regions
outside the event horizon of the Kerr spacetime. These forms
saturate the conditions (\ref{e:Om-infcon}) and
(\ref{e:psi-I-infcon}) at infinity. Inserting the expanded forms
into the stream equation, solutions can be derived order by order.
The solving procedure is as follows.

First, we need to choose the right zero-th order functions $\Om_0$,
$\psi_0$ and $I_0$, i.e., the conditions at infinity. But we only
need to know two of them, because the third one can be determined
via Eq.\ (\ref{e:I0exp}) (or (\ref{e:genI0con}) more generally) when
the two are given.

Second, we need to know the function forms $\Om(\psi)$ and $I(\psi)$
of $\psi$ (we can also take $\psi(\Om)$ and $I(\Om)$ as functions of
$\Om$). The function relations can be simply determined by the
zero-th order ones:
\begin{equation}\label{e:}
 \psi,\Om(\psi),I(\psi) \Longleftrightarrow
\psi_0,\Om_0(\psi_0),I_0(\psi_0),
\end{equation}
since the former will always lead to the latter as
$r\rightarrow\infty$. With the specific forms of the functions
$\Om(\psi)$ and $I(\psi)$, we can determine the values $\psi_+$,
$\Om_+$ and $I_+$ at the horizon by inserting the functions into the
Znajek regularity condition (\ref{e:Z+bd}). This is how the
conditions at the two boundaries are correlated. So the zero-th
order functions can be adjusted if the conditions at the horizon are
found to be inappropriate.

Finally, with all the functions and expanded forms inserted into the
stream equation, we can solve the equation order by order. The
obtained solutions should apply for rotating black holes with
general $a$.

In summary, the derived solutions in this method completely rely on
the choices of the conditions at the two boundaries. Given any two
of $\psi_0$, $\Om_0$ and $I_0$, the general function relations among
$\psi$, $\Om$ and $I$ can be determined by the zero-th order ones.
This further leads to the determinant of the condition at the
horizon. So, with appropriate conditions at both boundaries, a set
of solutions are defined.

Besides, there is an extra problem that needs to be classified: the
convergency at the horizon in the extreme limit. The coefficient of
the $n$-th order term of the derived solution should be order of
\begin{equation}\label{e:mn}
 \psi_{-n}\sim\mathcal{O}(m^n), \textrm{ }\textrm{ }\textrm{ }
(n\geq1)
\end{equation}
where
\begin{equation}\label{e:}
 m^n=\prod_{i=1}^{n}m_i, \textrm{ }\textrm{ }\textrm{ } m_i=(a,M).
\end{equation}
Thus, in the extreme limit $r_+=M=a$, each term of the expanded
forms (\ref{e:exp}) is order $\mathcal{O}(1)$ at the coincident
horizon. So every term is important close to the horizon in the
extreme case. We must check this convergency of the solution, which
is hard to do because we usually can not derive the full solution of
all orders. Fortunately, its convergency should be guaranteed by the
Znajek regularity condition at the horizon, since it applies for
arbitrary $a$.

\section{Solutions}
\label{sec:solutions}

As examples, we shall adopt the special boundary conditions obtained
in Section.\ (\ref{sec:spebc}) to make solutions in what follows.

\subsection{$\Om_0=\Om_P$}

As shown in \cite{Menon:2005va}, this case may correspond to the
split monopole because it is expanded to leading order of $a$ as
$\Om_P=a/(8M^2)+\cdots$ in the slow-rotating limit. Thus, we take
the zero-th order flux $\psi_0$ as that in the split monopole
solution (on the upper half hemisphere $0\leq\th<\pi/2$)
\begin{equation}\label{e:}
 \psi_0=\al(1-\cos\th),
\end{equation}
with $\al$ a constant. Then we have from Eq.\ (\ref{e:I0exp})
\begin{equation}\label{e:}
 I_0=-\frac{\al}{4\pi}\sin^2\th\Om_P.
\end{equation}

In terms of the relations between $\psi_0$ and $I_0(\psi_0)$,
$\Om_0(\psi_0)$, we can generalise them by assuming that the
relations apply for any $r$:
\begin{equation}\label{e:OmP-funrelation}
 \Om(\psi)=\frac{\al^2 a}{B(\psi)},
\textrm{ }\textrm{ }\textrm{ } I(\psi)=-\frac{\al
a\psi(2\al-\psi)}{4\pi B(\psi)},
\end{equation}
where $B(\psi)=\al^2(r_+^2+2Mr_+)+a^2(\al-\psi)^2$. Thus, we have
\begin{equation}\label{e:}
 II'=\frac{Mr_+\al^4a^2\psi(\al
-\psi)(2\al-\psi)}{2\pi^2B^3}.
\end{equation}

Inserting the functions $\Om(\psi)$ and $I(\psi)$ into the equation
(\ref{e:reGSeq}), we get an equation of $\psi$. The resulting
equation can be solved order by order by using the expanded forms
(\ref{e:exp}), in analogy to the Schwarzschild case shown in the
Appendix. Similarly, let us define
\begin{equation}\label{e:}
 L_\th^2\equiv\pa_\th^2+\left(2a\Om_P\sin2\th+\cot\th\right)
\pa_\th+6-8a\Om_P\cos^2\th-\frac{4}{\sin^2\th}.
\end{equation}

The vanishing of the coefficients of $r^0$ gives rise to the
equation about $\psi_0$, which is automatically saturated because it
is just the condition chosen at infinity. Comparing all the terms at
order $r^{-1}$ leads to the following equation about $\psi_{-1}$:
\begin{equation}\label{e:}
 L^2_\th\psi_{-1}=0.
\end{equation}
This equation can be solved by
\begin{equation}\label{e:}
 \psi_{-1}=\be a\sin^2\th,
\end{equation}
with $\be$ being an arbitrary dimensionless constant.

In order to make higher order calculations simpler, we may set the
free parameter $\be$ to be the special value $0$. Then the equation
about $\psi_{-2}$ can be obtained and simplified:
\begin{equation}\label{e:}
 (L^2_\th+2)\psi_{-2}=-8Mr_+\al a\cos\th
\sin^2\th\Om_P.
\end{equation}
A solution to the equation is
\begin{equation}\label{e:}
 \psi_{-2}=\frac{1}{2}\al a^2\cos\th\sin^2\th.
\end{equation}
Accurate to this order, the solution is somehow similar to the
slow-rotating solution in the large $r$ limit:
$A_\phi=C(1-\cos\th)+\mathcal{O}(a^2/M)(C\cos\th\sin^2\th)r^{-1}
+\cdots$, obtained in the perturbation approach
\cite{Blandford:1977ds}. The difference is in that the
next-to-leading order is at $r^{-2}$ for the former and it is at
$r^{-1}$ for the latter.

The equation of $\psi_{-3}$ is
\begin{equation}\label{e:}
 (L^2_\th+6)\psi_{-3}=2\al M\cos\th\left(3a\Om_P^{-1}
+4Mr_+-8Mr_+a\sin^2\th\Om_P\right).
\end{equation}
No analytical solution is found for the equation and so the
calculation procedure can not proceed.

Inserting the relations in Eq.\ (\ref{e:OmP-funrelation}) into the
the Znajek condition (\ref{e:Z+bd}), we can find that the condition
of $\psi$ at the horizon is the same as the one at infinity:
$\psi_+=\psi_0$, as we mentioned in the previous section. This means
that all higher order terms $\psi_{-n}$ ($1\leq n<\infty$) of a
legal solution $\psi$ must cancel out on the horizon, which is a
constraint of the Znajek regularity condition.

At boundaries, the solution satisfies $\Om_0=\Om_+\geq\om_+/2$,
where the equality occurs at $\th=0$. Generally, the solution up to
the second-order also satisfies
\begin{equation}
 \Om(r,\th)=\frac{a}{2Mr_++r_+^2+a^2\cos^2\th(1-\frac{1}{2}
a^2\sin^2\th r^{-2})^2}>\frac{1}{2}\om_+.
\end{equation}
This means that the magnetosphere in the valid regions is stable
\cite{Tomimatsu:2001wr,Pan:2015haa} against the screw instability
\cite{Li:2000mc}. Since the obtained solution is quite similar to
the split monopole perturbation solution at large $r$, other
properties about the solution will be not reconsidered here.

\subsection{$\Om_0=\Om_N$}

\subsubsection{The expanded solution}

As stated previously, $\Om=\Om_N$ is the MD exact solution of the
force-free magnetosphere on general rotating black holes. But this
$r$-independent solution is unrealistic. It is interesting to
investigate the situation by extending the solution to the
$r$-dependent case through the expansion approach given above.

We take $\Om_N$ as the initial value at the boundary to derive the
$r$-dependent solution. Obviously, $\Om_0=\Om_N$ is singular at
poles $\th=0,\pi/2$. Thus, we demand $\psi_0$ to be non-singular by
taking the simple form:
\begin{eqnarray}\label{e:psi0Om0}
 \psi_0=c\Om_0^{-k}+d, & (c>0, & k>0)
\end{eqnarray}
where $c$, $d$ and $k$ are constants. By choosing positive sign in
Eq.\ (\ref{e:genI0con}) instead, we have
\begin{equation}\label{e:}
 I_0=\frac{kc}{2\pi}\cos\th\Om_0^{1-k},
\end{equation}
since $\Om_0=1/(a\sin^2\th)$ is already faster than $\om_+$.

In terms of the relations among the zero-th order functions, we can
get the functional relations at general $r$:
\begin{equation}\label{e:OmN-psi-Om}
 \psi(\Om)=c\Om^{-k}+d,
\end{equation}
\begin{equation}\label{e:OmN-I-Om}
 I(\Om)=\frac{kc}{2\pi}\sqrt{1-(a\Om)^{-1}}\Om^{1-k},
\end{equation}
which lead to
\begin{equation}\label{e:}
 \frac{16\pi^2}{kc}II'\Om^{2+k}=4\left(k
-1\right)\Om^{4}-\frac{2(2k-1)}{a}\Om^{3}.
\end{equation}

It is convenient for later calculations to redefine
\begin{equation}\label{e:}
 \widetilde{\Om}\equiv\frac{\Om}{\Om_N}
\textrm{ }\textrm{ }\textrm{ with }\textrm{ }\textrm{ }
\widetilde{\Om}_{-n}(\th)\equiv\frac{\Om_{-n}(\th)}{\Om_N},
\end{equation}
With the above relations, then the stream equation can be expressed
as
\begin{eqnarray}\label{e:OmNGseq}
 \rho^2\widetilde{\Om}[A\widetilde{\Om}(\widetilde{\Om}-
2a\sin^2\th\om)-a^2\sin^2\th(\rho^2-2Mr)](\triangle\pa_r^2
\widetilde{\Om}+\pa_\th^2\widetilde{\Om})
\nonumber \\
-\rho^2[A\widetilde{\Om}(k\widetilde{\Om}-(2k+1)a\sin^2\th\om)
-(k+1)a^2\sin^2\th(\rho^2-2Mr)]
[\triangle(\pa_r\widetilde{\Om})^2+(\pa_\th\widetilde{\Om})^2]
\nonumber \\
+2\triangle\widetilde{\Om}[r\rho^4\widetilde{\Om}^2-
Ma^2\sin^2\th(r^2-a^2\cos^2\th)(\widetilde{\Om}-1)^2]
\pa_r\widetilde{\Om}
\nonumber \\
+\cot\th\widetilde{\Om}\{(4k-1)\rho^2[A\widetilde{\Om}
(\widetilde{\Om}-2a\sin^2\th\om)-a^2\sin^2\th(\rho^2-2Mr)]
\nonumber \\
+2a^2\sin^2\th A[\widetilde{\Om}(\widetilde{\Om}-2a\sin^2\th\om)
+a\sin^2\th\om]-2\rho^2(r^2+a^2)^2\widetilde{\Om}^2\}\pa_\th
\widetilde{\Om}
\\
+2[2(k-1)\rho^2(A-a^2(\rho^2+2Mr))+2Mr(r^2+a^2)(r^2-a^2\cos^2\th)
+\rho^4\triangle]\widetilde{\Om}^4
\nonumber \\
+2[(2k-1)\rho^2(4Mra^2\cos^2\th-\rho^4)-4Mra^2\sin^2\th(r^2
-a^2\cos^2\th)]\widetilde{\Om}^3
\nonumber \\
+2[2(k-1)\rho^2a^2\cos^2\th(\rho^2-2Mr)-a^2\sin^2\th(\rho^4
-2Mr(r^2-a^2\cos^2\th))]\widetilde{\Om}^2=0. \nonumber
\end{eqnarray}

Inserting the expanded form of $\widetilde{\Om}$ into the above
equation, we can get an expanded equation. The vanishing of the
coefficients of $r^{6-n}$ gives rise to the equation about
$\widetilde{\Om}_{-n}$ ($n\geq1$), which can be formally expressed
as:
\begin{equation}\label{e:expOmNGseq}
 [L_\th^2+n(n-1)]\widetilde{\Om}_{-n}
=F_{-n}(\widetilde{\Om}_0,\widetilde{\Om}_{-1},\cdots,
\widetilde{\Om}_{-(n-1)}), \textrm{ }\textrm{ }\textrm{ } (n\geq1)
\end{equation}
where
\begin{equation}\label{e:}
 L_\th^2=\pa_\th^2+(4k-3)\cot\th\pa_\th+2(2k-1).
\end{equation}
The functions $F_{-n}$ at order $-n$ are some functions of
$\widetilde{\Om}_{-i}$ with $0\leq i\leq n-1$.

The fact that $\Om=\Om_N$ is an exact solution to the full stream
equation means that we can always have for all $n\geq1$
\begin{equation}\label{e:}
 [L_\th^2+n(n-1)]\widetilde{\Om}_{-n}=0 \textrm{ }\textrm{ }
\textrm{ with }\textrm{ }\textrm{ } \widetilde{\Om}_{-n}=0.
\end{equation}
In what follows we shall consider more general ($r$-dependent)
solutions other than this trivial case.

With $\Om_0=\Om_N$, the vanishing of the terms at order $r^{6}$ is
automatically satisfied, as expected. For the order $r^5$, the
obtained equation is
\begin{equation}\label{e:Om_Npsi-1}
 L_\th^2\widetilde{\Om}_{-1}=0.
\end{equation}
The equation has the simple solution
\begin{equation}\label{e:psi-1sol1}
 \widetilde{\Om}_{-1}=-2\al a\cos\th,
\end{equation}
where $\al$ is an arbitrary dimensionless constant. We assume that
the solution applies to the upper hemisphere $\th\in[0,\pi/2]$ since
it is asymmetric about the equatorial plane.

The equation (\ref{e:Om_Npsi-1}) has the second kind of solution,
which is symmetric. The solution generally can be expressed in terms
of the hypergeometric functions. But we can have their explicit
forms when $4k-3$ is an odd number. For example, the solution for
$k=3/2$ is
\begin{equation}\label{e:psi-1sol2}
 \widetilde{\Om}_{-1}=-\be a\left[1-\frac{1}{2}\cot^2\th+
\frac{3}{4}\cos\th\ln\frac{1-\cos\th}{1+\cos\th}\right],
\end{equation}
where $\be$ is a dimensionless constant. This solution is symmetric
under $\cos\th\rightarrow-\cos\th$. But the solution forms closed
magnetic field lines, which is excluded for a force-free
magnetosphere \cite{MacDonald:1982zz,Gralla:2014yja}. So this
solution is abandoned.

At order $r^4$, the resulting equation about $\Om_{-2}$ can be
simply reduced by inserting the solution (\ref{e:psi-1sol1})
\begin{equation}\label{e:}
 (L_\th^2+2)\widetilde{\Om}_{-2}=4k\al^2a^2.
\end{equation}
A simple solution to this equation is
\begin{equation}\label{e:psi-2}
 \widetilde{\Om}_{-2}=\al^2a^2.
\end{equation}
The equation of $\widetilde{\Om}_{-2}$ by inserting the second
solution (\ref{e:psi-1sol2}) is hard to be solved and is not
considered.

The vanishing of the coefficients of $r^3$ leads to the equation
about $\Om_{-3}$, which can be reduced to
\begin{equation}\label{e:}
 (L_\th^2+6)\widetilde{\Om}_{-3}=
4\al^2Ma^2(1-2k\cos^2\th)+4\al a^3\cos\th[3+2(1-2k)\cos^2\th].
\end{equation}
When $k\neq3/2$, a solution to this equation is
\begin{equation}\label{e:psi-3}
 \widetilde{\Om}_{-3}=2\al a^3\cos^3\th+\frac{\al^2Ma^2}{2k-3}
\left(4k\cos^2\th-\frac{3}{k+1}\right).
\end{equation}
The solution at the critical value $k=3/2$ is not found.

The solutions at higher orders are hard to derive due to the
involvement of many more terms. But we can make some simple analysis
based on the equation (\ref{e:OmNGseq}) and the derived solutions
above. For higher orders, we can find that the function $F_{-n}$ for
each $n$ in Eq.\ (\ref{e:expOmNGseq}) should be some polynomial of
$\cos\th$:
\begin{equation}\label{e:}
 F_{-n}(\th)=\sum_i^nf_i(\al,k)\cos^i\th,
\end{equation}
where $f_i$ are coefficients and are of order $m^n$, as pointed out
in Eq.\ (\ref{e:mn}). So the equation (\ref{e:expOmNGseq}) with the
form of $F_{-n}$ should be solvable except that abnormal situations
emerge. This implies that an exact solution may be eventually
obtained or guessed by following the procedure if we could
successfully handle all the terms to higher enough orders.

At the moment, the above solution up to the first few orders should
be valid for asymptotical regions far away from the horizon. The
solution is consistent with our near-horizon solution for
near-extreme black holes \cite{Li:2014bta}. It can be checked that
the solution forms open magnetic field lines, which may be separated
by a current sheet on the equatorial plane, just like the split
monopole solution.

\subsubsection{Analysis of the solution}

Our solution generslises the MD exact solution $\Om=\Om_N$
\cite{Menon:2005va,Menon:2005mg} to the ($r,\th$)-dependent case.
The MD exact solution is taken as an initial condition at both
boundaries and is recovered from the generalised solution as the
parameter $\al=0$. The exact solution has difficulties to describe a
realistic magnetosphere since its four-current and the
electromagnetic field are both null. Here we examine the situation
for our generalised solution.

Before doing that, we first determine the conditions that the
quantities from the solution are not singular on the poles $\th=0$,
which are summarised in Table \ref{t:t1}. For $k\geq3/2$, all the
quantities on the table (as well as the electromagnetical fields)
are non-singular on the poles.

\begin{table}
\begin{center}
\begin{tabular}{|c|c|c|c|c|}
\hline
$\psi$ & $\mathcal{L}^\th$ & $I$, $\mathcal{L}^r$ & $\mathcal{E}^\th$ & $\mathcal{E}^r$ \\
\hline
$k\geq0$ & $k\geq\frac{3}{4}$ & $k\geq1$ & $k\geq\frac{5}{4}$ & $k\geq\frac{3}{2}$ \\
\hline
\end{tabular}
\caption{\label{t:t1} The conditions of $k$ for the corresponding
quantities to be non-singular on the rotation axis.}
\end{center}
\end{table}

\begin{figure*}
\centering
\includegraphics[angle=0,scale=0.5]{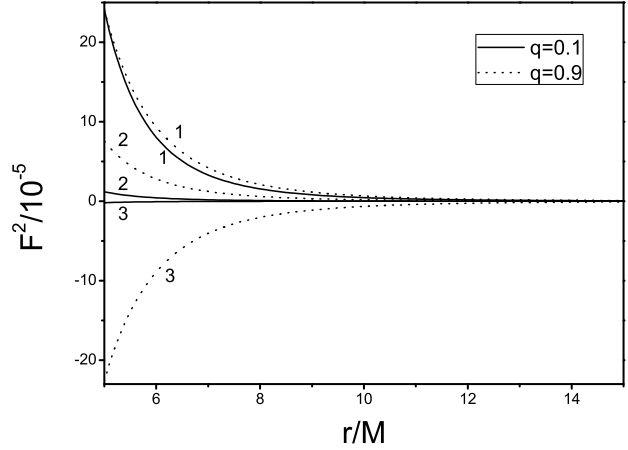}
\caption{\label{f:f1} Illustrations of the invariant $F^2$ at
different radial distances $r$ and different poloidal angles
$\th=q\pi/2$. The parameter $c=1$ and the spin parameter $a=0.8M$.
The line groups: (1) $\al=1$ and $k=1.3$; (2) $\al=0.1$ and
$k=1.49$; (3) $\al=1$ and $k=2$.}
\end{figure*}

The existence of magnetosphere in all frames requires it should be
magnetically dominated, i.e., the following invariant to be
\begin{equation}\label{e:J2}
 F^2=2(\textbf{B}^2-\textbf{E}^2)>0.
\end{equation}
Inserting the solution into the expressions, we can get the
invariant. We find that its sign is strongly affected by the
parameter $k$ but not sensitive to other parameters like $\al$ and
$a$. As shown in Fig.\ \ref{f:f1}, the invariant $F^2$ is positive
for $k<3/2$ while negative for $k>3/2$ (the case $k=3/2$ can not be
judged since the solution is not available here). This indicates
that the magnetic fields can get dominated only when (part of) the
quantities are singular on the poles. As expected, the values all
asymptotically approach 0 at large $r$ as the ($r,\th$)-dependent
solution recovers the MD exact solution at the far boundary.

Whether the four-current $J^\mu$ is timelike, lightlike or spacelike
can be determined by judging $J^2=J_\mu J^\mu$ (contracted by the
four-dimensional metric (\ref{e:Kerr})) to be negative, null, or
positive, respectively. The contracted current is related to the
charge and current densities measured in ZAMOs via
\begin{equation}\label{e:J2}
 J^2=-\rho_e^2+\textbf{j}\cdot\textbf{j},
\end{equation}
with their components satisfying
\begin{equation}\label{e:currentrel1}
 \frac{1}{\La}\rho_e=J^0, \textrm{ }\textrm{ }\textrm{ }
j^r=J^r,
\end{equation}
\begin{equation}\label{e:currentrel1}
 j^\th=J^\th, \textrm{ }\textrm{ }\textrm{ }
j^\phi=-\om J^0+J^\phi.
\end{equation}
The three-current $\textbf{j}$ is contracted by the metric
(\ref{e:absolutesp}) of the absolute space.

\begin{figure*}
\centering
\includegraphics[angle=0,scale=0.39]{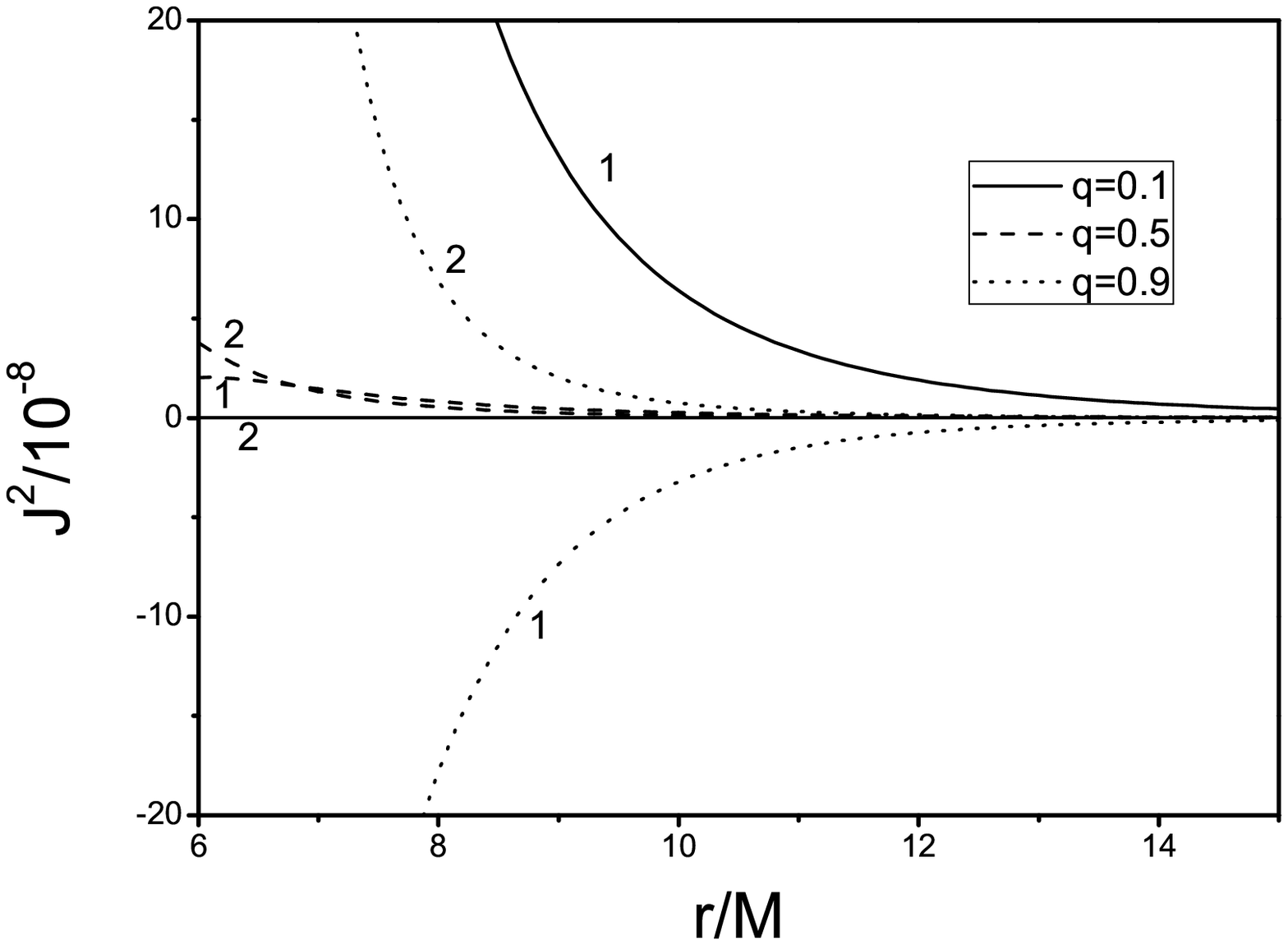}
\includegraphics[angle=0,scale=0.39]{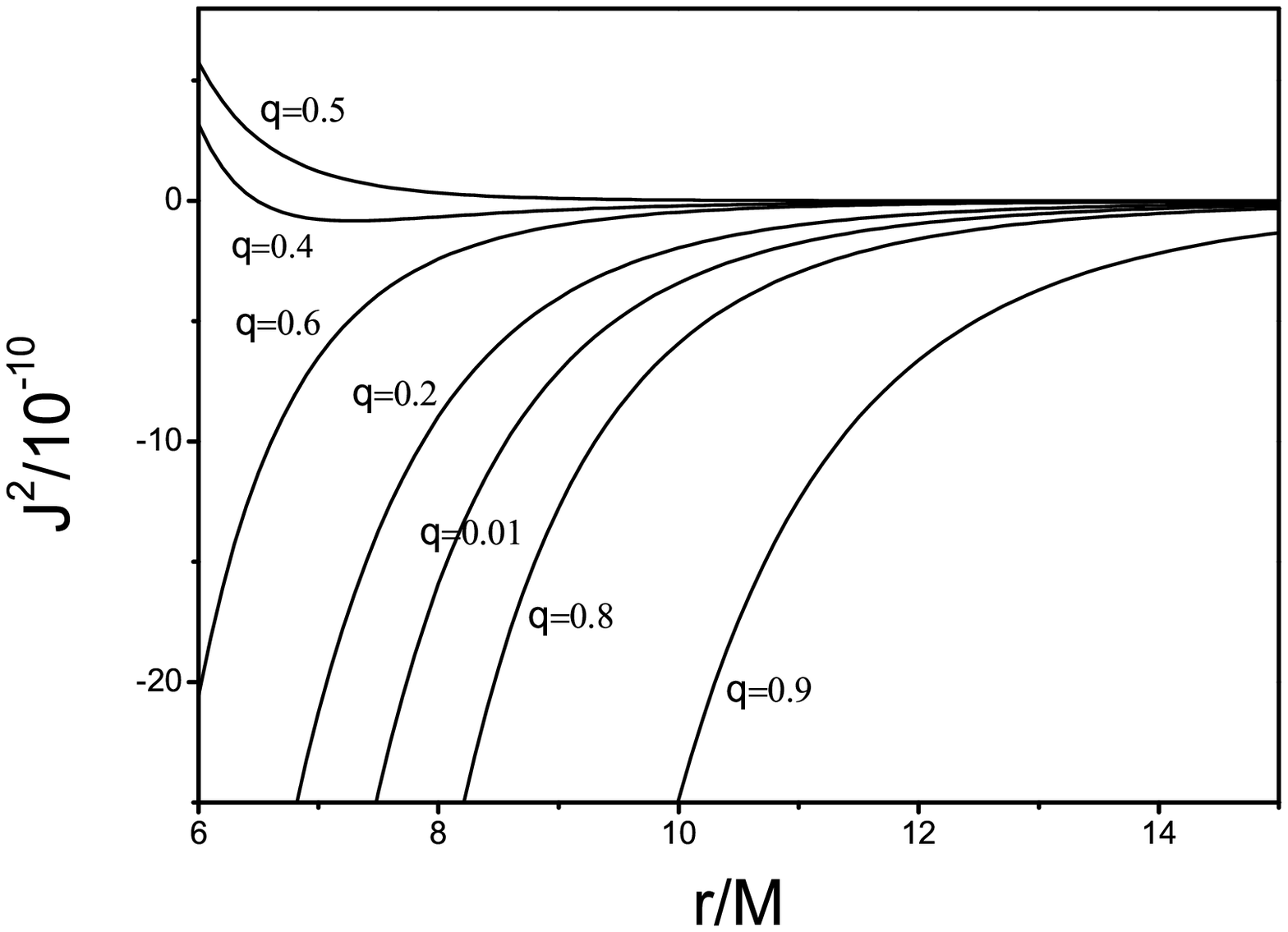}
\caption{\label{f:f2} Illustrations of $J^2$ at different distances
$r$ and different angles $\th=q\pi/2$. The parameters are chosen to
be: $c=1$ and $a=0.8M$. \textit{Left panel}: (1) $\al=1$ and
$k=1.3$; (2) $\al=1$ and $k=3$. \textit{Right panel}: $\al=0.1$ and
$k=1.49$.}
\end{figure*}

The sign of $J^2$ is also sensitive to $k$, as shown in Fig.\
\ref{f:f2}. For $k>3/2$, the values of $J^2$ are almost all positive
at all angles $\th$. For $k<3/2$, they are not always positive and
are negative for larger $\th$, i.e., near the equatorial plane. The
only case that its values are mostly negative happens when
$k\rightarrow3/2$ from the $k<3/2$ side. The case $k=1.49$ (to
regularise $\widetilde{\Om}_3$ to be not too large, we adopt a small
$\al=0.1$) is shown on the right panel of Fig.\ \ref{f:f2}. It can
be seen that the values of $J^2$ grow with $\th$ increasing from
negative values at the small angle $\th=0.01\pi/2$, and turn to be
slightly positive at around $\th=\pi/4$. Then they turn back to be
negative again for larger angles. It can be checked that the values
of $J^2$ are also negative for angles smaller than $\th=0.01\pi/2$.
But they all will tend to be null: $J^2=0$, at exactly $\th=0$.

\section{Summary}
\label{sec:summary}

In this work, we adopt a new expansion method to explore analytical
solutions of force-free magnetospheres on black holes with arbitrary
spin parameter. The functions and stream equation are expanded in
series of the radial distance in terms of the boundary conditions at
the event horizon and at spatial infinity. With the conditions at
the two boundaries chosen, a set of solutions can be defined and
solved order by order.

In terms of the regular conditions at both boundaries, the two
asymptotical solutions found by MD in
\cite{Menon:2005va,Menon:2005mg} are identified as the solutions
that have the same conditions at the two boundaries. By taking them
as initial conditions at the boundaries, we derived the
corresponding expanded solutions to higher orders. The first one
corresponds to the split monopole solution obtained in the
perturbation approach when we take the $a\rightarrow0$ limit. It is
found to have similar asymptotical profile to the latter at large
$r$, though not with the same $r$ dependence.

The second solution can be viewed as an extension of the
$r$-independent MD exact solution to the ($r,\th$)-dependent case.
With an appropriate choice of the relation between $\psi$ and $\Om$
at the far boundary, we find that the expanded stream equation
should be solvable at each order. So an exact solution (probably
with a closed form) is hopefully derived or guessed if we could
calculate to all or higher enough orders, though we only get the
expanded solution up to first few orders in this work.

Based on the obtained solution, we show that the extended solution
can (partially) avoid the problems of the $r$-independent MD
solution: the four-current and the electromagnetic field are both
null. When the parameter $k$ tends to the critical value $3/2$ from
the $k<3/2$ side, our solution leads to a force-free magnetosphere
which is magnetically dominated with timelike current in most
directions $\th$. The current gets slightly spacelike at around
$\th=\pi/4$ and lightlike at exactly $\th=0$. A difficulty for the
solution with $k$ less than and close to $3/2$ is that the energy
extraction (integration of $\mathcal{E}^r$) highly converges along
the rotation axis in a singular way. Similar singular behaviours
also exist in the relieving method
\cite{Menon:2011zu,Menon:2014vqa}. But, in our case, the singular
mode is very slight for $k\rightarrow3/2$. Nevertheless, we may
still have to exclude the $\th=0$ direction or assume that the
force-free condition is violated by dense plasma in this region.

As we can see, the solution with $k=3/2$ is an interesting case, but
is not found in this work, which is left for further study. We also
need to derive the expanded solution to higher orders and to check
whether the results from the present solution still hold (or even
improve). Moreover, more varieties of the relation between $\psi_0$
and $\Om_0$ other than (\ref{e:psi0Om0}) are under consideration to
find saturate results.

\section*{Acknowledgements\markboth{Acknowledgements}{Acknowledgements}}

This work is partially supported by the National Natural Science
Foundation of China 11573060, 11661161010.

\appendix

\section*{Appendix}

\setcounter{equation}{0}

\section{Solutions on a Schwarzschild black hole}
A detailed discussion of exact solutions of magnetospheres on
Schwarzschild black holes haven been made in \cite{Ghosh:1999in}.
Here, we use the expansion method in the text to rederive the
solutions. The Schwarzschild metric is
\begin{equation}\label{e:Sch}
 ds^2=-\left(1-\frac{r_0}{r}\right)dt^2+\left(1-\frac{r_0}{r}
\right)^{-1}dr^2+r^2d\Om^2,
\end{equation}
where the horizon is located at $r_0=2M$.

In the non-rotating case, the stream equation reduces to
\begin{equation}\label{e:Maxeq-Sch}
 x^2\pa_x\left[\left(1-x^{-1}\right)\pa_x\psi\right]
+L_\th^2\psi=0,
\end{equation}
where
\begin{equation}\label{e:}
 x\equiv\frac{r}{r_0},
\end{equation}
\begin{equation}\label{e:}
 L_\th^2=\sin\th\pa_\th\left(\sin^{-1}\th\pa_\th\right)
=\pa_\th^2-\cot\th\pa_\th.
\end{equation}
Here, we take $\psi$ to be dimensionless to simplify the notation.
This equation is essentially the Maxwell's equation on Schwarzschild
in the absence of sources, i.e., $\rho_e=\textbf{j}=0$.

So the force-free condition (\ref{e:ffcon}) is trivial here and does
not really provide any extra constraint. We need to find alternative
boundary conditions. Let us adopt the ansatz of a general solution:
\begin{equation}\label{e:}
 \psi(x,\th)=\psi_1(\th)x+\psi_*\ln x+\psi_0(\th)+\psi_{-1}(\th)x^{-1}
+\psi_{-2}(\th)x^{-2}+\cdots.
\end{equation}
We choose this ansatz to guarantee that the electromagnetic fields
vanish at $x\rightarrow\infty$.

Inserting the expanded form into the equation and comparing the
coefficients of each order of $x$, we get the following equations:
\begin{equation}\label{e:psi1}
 L_\th^2\psi_1=0,
\end{equation}
\begin{equation}\label{e:psistar}
 L_\th^2\psi_*=0,
\end{equation}
\begin{equation}\label{e:psi0}
 L_\th^2\psi_0=\psi_*-\psi_1,
\end{equation}
\begin{equation}\label{e:psi-1}
 (L_\th^2+2)\psi_{-1}=-2\psi_*,
\end{equation}
\begin{equation}\label{e:psi-n}
 [L_\th^2+n(n+1)]\psi_{-n}=(n^2-1)\psi_{-(n-1)}.
\textrm{ }\textrm{ }\textrm{ } (n\geq2)
\end{equation}

(I) Non-separable solutions

The first four equations (\ref{e:psi1})-(\ref{e:psi-1}) are closed
and complete to give exact solutions. We first set the coefficients
$\psi_{-n}=0$ ($n\geq1$) so that $\psi_*=0$ since they are always
solutions. Then from Eq.\ (\ref{e:psi1}), a solution of $\psi_1$ can
be written as the form
\begin{equation}\label{e:}
 \psi_1
=1-\cos\th,
\end{equation}
With it, the equation (\ref{e:psi0}) can be expressed as
\begin{equation}\label{e:}
 \pa_y^2\psi_0=-\frac{1}{y},
\end{equation}
where $y=1+\cos\th$. A general solution to this is
\begin{equation}\label{e:}
 \psi_0=
\al+\be\cos\th-(1+\cos\th)\ln(1+\cos\th),
\end{equation}
where $\al$ and $\be$ are constants. When $-\al=\be=1$, the full
solution is
\begin{equation}\label{e:}
 \psi=(x-1)(1-\cos\th)-(1+\cos\th)\ln(1+\cos\th),
\end{equation}
which is exactly the non-separable solution \cite{Ghosh:1999in}.

(II) Separable solutions

(1) Zeroth-order solution

From Eqs.\ (\ref{e:psi1})-(\ref{e:psi0}), we can impose the general
solution
\begin{eqnarray}\label{e:}
 \psi_1=\psi_*=b+c\cos\th, &
\psi_0=d\psi_1+\al\cos\th+\be,
\end{eqnarray}
where $b$, $c$, $d$ and $e$ are constants. So we have
\begin{equation}\label{e:}
 \psi_{-n}=-\frac{1}{n}\psi_1. \textrm{ }\textrm{ }\textrm{ } (n\geq1)
\end{equation}
Adopting the relation $\ln(1-z)=-\sum_{n=1}^{\infty}z^n/n$, we can
express the solution as
\begin{equation}\label{e:}
 \psi=\al\cos\th+\be+(b+c\cos\th)[d+x+\ln(x-1)].
\end{equation}
This is the lowest order separable solution with $m=0$ given in
\cite{Ghosh:1999in}. The case $b=c=0$ is the (split) monopole
solution.

(2) First-order solution

From the first three equations (\ref{e:psi1})-(\ref{e:psi0}), we
consider the case:
\begin{equation}\label{e:}
 \psi_1=\psi_*=0,
\end{equation}
and
\begin{equation}\label{e:}
 \psi_0=\al\cos\th+\be.
\end{equation}
Then Eq.\ (\ref{e:psi-1}) becomes $L_\th^2\psi_{-1}=-2\psi_{-1}$. So
the general solution of $\psi_{-1}$ can be
\begin{equation}\label{e:}
 \psi_{-1}=g\sin^2\th,
\end{equation}
where $g$ is an arbitrary constant. Thus, we can have generically
from Eq.\ (\ref{e:psi-n}):
\begin{equation}\label{e:}
 \psi_{-n}=\frac{3g}{n+2}\sin^2\th.
\textrm{ }\textrm{ }\textrm{ } (n\geq2)
\end{equation}
By using the expansion expression of $\ln(1-z)$, we can express the
full solution as
\begin{equation}\label{e:}
 \psi=\al\cos\th+\be-3g\sin^2\th\left[\frac{1}{2}+x+x^2
\ln\left(1-\frac{1}{x}\right)\right].
\end{equation}
The solution with $\al=\be=0$ is clearly the separable solution at
the order $m=1$ given in \cite{Ghosh:1999in}.

(3) Higher order solutions

If we consider the case $\psi_1=\psi_*=\psi_0=\psi_{-1}=0$, then
Eq.\ (\ref{e:psi-n}) becomes $L_\th^2\psi_{-2}=-6\psi_{-2}$. It
solves as $\psi_{-2}=3h\cos\th\sin^2\th$. We can then insert the
solution into the general $\psi_{-n}$. Following the same approach
above, we can derive the separable solution at the $m=3$ order.
Similarly, we can derive all higher order separable solutions.



\bibliographystyle{JHEP}
\bibliography{b}

\providecommand{\href}[2]{#2}\begingroup\raggedright\begin{thebibliography}{10}

\bibitem{Blandford:1977ds}
R.~Blandford and R.~Znajek, \emph{{Electromagnetic extractions of energy from
  Kerr black holes}}, {\emph{Mon.Not.Roy.Astron.Soc.} {\bfseries 179} (1977)
  433--456}.

\bibitem{Znajek:1977unknown}
R.~Znajek, \emph{{Black hole electrodynamics and the Carter tetrad}},
  {\emph{Mon.Not.Roy.Astron.Soc.} {\bfseries 179} (1977) 457--472}.

\bibitem{Tanabe:2008wm}
K.~Tanabe and S.~Nagataki, \emph{{Extended monopole solution of the
  Blandford-Znajek mechanism: Higher order terms for a Kerr parameter}},
  \href{https://doi.org/10.1103/PhysRevD.78.024004}{\emph{Phys.Rev.} {\bfseries
  D78} (2008) 024004}, [\href{https://arxiv.org/abs/0802.0908}{{\ttfamily
  0802.0908}}].

\bibitem{Pan:2015haa}
Z.~Pan and C.~Yu, \emph{{Fourth-order split monopole perturbation solutions to
  the Blandford-Znajek mechanism}},
  \href{https://doi.org/10.1103/PhysRevD.91.064067}{\emph{Phys. Rev.}
  {\bfseries D91} (2015) 064067},
  [\href{https://arxiv.org/abs/1503.05248}{{\ttfamily 1503.05248}}].

\bibitem{Menon:2005va}
G.~Menon and C.~D. Dermer, \emph{{Analytic solutions to the constraint equation
  for a force-free magnetosphere around a kerr black hole}},
  \href{https://doi.org/10.1086/497631}{\emph{Astrophys.J.} {\bfseries 635}
  (2005) 1197--1202}, [\href{https://arxiv.org/abs/astro-ph/0509130}{{\ttfamily
  astro-ph/0509130}}].

\bibitem{Menon:2005mg}
G.~Menon and C.~D. Dermer, \emph{{A class of exact solution to the
  blandford-znajek process}},
  \href{https://doi.org/10.1007/s10714-007-0418-2}{\emph{Gen.Rel.Grav.}
  {\bfseries 39} (2007) 785--794},
  [\href{https://arxiv.org/abs/astro-ph/0511661}{{\ttfamily
  astro-ph/0511661}}].

\bibitem{Brennan:2013jla}
T.~D. Brennan, S.~E. Gralla and T.~Jacobson, \emph{{Exact Solutions to
  Force-Free Electrodynamics in Black Hole Backgrounds}},
  \href{https://doi.org/10.1088/0264-9381/30/19/195012}{\emph{Class.Quant.Grav%
.} {\bfseries 30} (2013) 195012},
  [\href{https://arxiv.org/abs/1305.6890}{{\ttfamily 1305.6890}}].

\bibitem{Menon:2011zu}
G.~Menon and C.~D. Dermer, \emph{{Jet Formation in the magnetospheres of
  supermassive black holes: analytic solutions describing energy loss through
  Blandford-Znajek processes}},
  \href{https://doi.org/10.1111/j.1365-2966.2011.19327.x}{\emph{Mon. Not. Roy.
  Astron. Soc.} {\bfseries 417} (2011) 1098},
  [\href{https://arxiv.org/abs/1105.4139}{{\ttfamily 1105.4139}}].

\bibitem{Menon:2014vqa}
G.~Menon and C.~Dermer, \emph{{Local, Non-Geodesic, Timelike Currents in the
  Force-Free Magnetosphere of a Kerr Black Hole}},
  \href{https://doi.org/10.1007/s10714-015-1896-2}{\emph{Gen. Rel. Grav.}
  {\bfseries 47} (2015) 52}, [\href{https://arxiv.org/abs/1408.3514}{{\ttfamily
  1408.3514}}].

\bibitem{Lupsasca:2014pfa}
A.~Lupsasca, M.~J. Rodriguez and A.~Strominger, \emph{{Force-Free
  Electrodynamics around Extreme Kerr Black Holes}},
  \href{https://doi.org/10.1007/JHEP12(2014)185}{\emph{JHEP} {\bfseries 1412}
  (2014) 185}, [\href{https://arxiv.org/abs/1406.4133}{{\ttfamily 1406.4133}}].

\bibitem{Lupsasca:2014hua}
A.~Lupsasca and M.~J. Rodriguez, \emph{{Exact Solutions for Extreme Black Hole
  Magnetospheres}}, \href{https://doi.org/10.1007/JHEP07(2015)090}{\emph{JHEP}
  {\bfseries 07} (2015) 090},
  [\href{https://arxiv.org/abs/1412.4124}{{\ttfamily 1412.4124}}].

\bibitem{Li:2014bta}
H.~Li, C.~Yu, J.~Wang and Z.~Xu, \emph{{Force-free magnetosphere on
  near-horizon geometry of near-extreme Kerr black holes}},
  \href{https://doi.org/10.1103/PhysRevD.92.023009}{\emph{Phys. Rev.}
  {\bfseries D92} (2015) 023009},
  [\href{https://arxiv.org/abs/1403.6959}{{\ttfamily 1403.6959}}].

\bibitem{Zhang:2014pla}
F.~Zhang, H.~Yang and L.~Lehner, \emph{{Towards an understanding of the
  force-free magnetosphere of rapidly spinning black holes}},
  \href{https://doi.org/10.1103/PhysRevD.90.124009}{\emph{Phys. Rev.}
  {\bfseries D90} (2014) 124009},
  [\href{https://arxiv.org/abs/1409.0345}{{\ttfamily 1409.0345}}].

\bibitem{Compere:2015pja}
G.~Comp¨¨re and R.~Oliveri, \emph{{Near-horizon Extreme Kerr Magnetospheres}},
  \href{https://doi.org/Phys. Rev.D93,no.6,069906(2016)}{\emph{Phys. Rev.}
  {\bfseries D93} (2016) 024035},
  [\href{https://arxiv.org/abs/1509.07637}{{\ttfamily 1509.07637}}].

\bibitem{MacDonald:1982zz}
D.~MacDonald and K.~S. Thorne, \emph{{Black-hole electrodynamics - an
  absolute-space/universal-time formulation}}, {\emph{Mon. Not. Roy. Astron.
  Soc.} {\bfseries 198} (1982) 345--383}.

\bibitem{Nathanail:2014aua}
A.~Nathanail and I.~Contopoulos, \emph{{Black Hole Magnetospheres}},
  \href{https://doi.org/10.1088/0004-637X/788/2/186}{\emph{Astrophys. J.}
  {\bfseries 788} (2014) 186},
  [\href{https://arxiv.org/abs/1404.0549}{{\ttfamily 1404.0549}}].

\bibitem{Ruiz:2012te}
M.~Ruiz, C.~Palenzuela, F.~Galeazzi and C.~Bona, \emph{{The Role of the
  ergosphere in the Blandford-Znajek process}},
  \href{https://doi.org/10.1111/j.1365-2966.2012.20950.x}{\emph{Mon. Not. Roy.
  Astron. Soc.} {\bfseries 423} (2012) 1300--1308},
  [\href{https://arxiv.org/abs/1203.4125}{{\ttfamily 1203.4125}}].

\bibitem{Tomimatsu:2001wr}
A.~Tomimatsu, T.~Matsuoka and M.~Takahashi, \emph{{Screw instability in black
  hole magnetospheres and a stabilizing effect of field line rotation}},
  \href{https://doi.org/10.1103/PhysRevD.64.123003}{\emph{Phys. Rev.}
  {\bfseries D64} (2001) 123003},
  [\href{https://arxiv.org/abs/astro-ph/0108511}{{\ttfamily
  astro-ph/0108511}}].

\bibitem{Li:2000mc}
L.-X. Li, \emph{{Screw instability and blandford-znajek mechanism}},
  \href{https://doi.org/10.1086/312538}{\emph{Astrophys. J.} {\bfseries 531}
  (2000) L111}, [\href{https://arxiv.org/abs/astro-ph/0001420}{{\ttfamily
  astro-ph/0001420}}].

\bibitem{Gralla:2014yja}
S.~E. Gralla and T.~Jacobson, \emph{{Spacetime approach to force-free
  magnetospheres}}, \href{https://doi.org/10.1093/mnras/stu1690}{\emph{Mon.
  Not. Roy. Astron. Soc.} {\bfseries 445} (2014) 2500--2534},
  [\href{https://arxiv.org/abs/1401.6159}{{\ttfamily 1401.6159}}].

\bibitem{Ghosh:1999in}
P.~Ghosh, \emph{{The Structure of black hole magnetospheres. 1. Schwarzschild
  black holes}},
  \href{https://doi.org/10.1046/j.1365-8711.2000.03410.x}{\emph{Mon. Not. Roy.
  Astron. Soc.} {\bfseries 315} (2000) 89},
  [\href{https://arxiv.org/abs/astro-ph/9907427}{{\ttfamily
  astro-ph/9907427}}].

\end{thebibliography}\endgroup

\end{document}